\documentclass[twoside,fleqn]{article}
\usepackage{espcrc2}
\usepackage{graphicx}

\title{
{\Large  Rho Meson Properties 
}}
\author{J.J. Sanz-Cillero and A. Pich
\address{
       {\em Instituto de F\'\i sica Corpuscular, Universitat de Val\`encia,} \\
       {\em Apartat Correus 22085, E-46071 Val\`encia, Spain}}
\vspace*{-0.5cm}{\sf \small \rightline{IFIC/03-44,  FTUV/03-1014}}  
       }

\begin{document}

\newcommand{\xs}{\mbox{$x(\sigma)$}}
\newcommand{\vs}{\vbox{\vskip 1cm plus .3cm minus .3cm}}
\def\question#1{{{\marginpar{\small \sc #1}}}}
\newcommand{\pom}{ $I\hspace{-1.6mm}P$}
\newcommand{\pome}{ I\hspace{-1.6mm}P}
\newcommand{\bean}{\begin{eqnarray*}}
\newcommand{\eean}{\end{eqnarray*}}
\newcommand{\kkpi}{\mbox{$K^{0}_{S} K^{\pm} \pi^{\mp}$} }
\newcommand{\etapipi}{\mbox{$\eta \pi \pi$} }
\newcommand{\kskbar}{\mbox{$K^* \overline K$} }
\newcommand{\gapproxeq}{\lower
.7ex\hbox{$\;\stackrel{\textstyle >}{\sim}\;$}}
\newcommand{\lapproxeq}{\lower
.7ex\hbox{$\;\stackrel{\textstyle <}{\sim}\;$}}
\newcommand{\pipipipi}{\mbox{$\pi^+\pi^-\pi^+\pi^-$ }}
\newcommand{\qbar}{\mbox{$\mathrm{\overline{q}}$} }
\newcommand{\JPC}{\mbox{J$\mathrm{^{PC}}$} }
\newcommand{\ubar}{\mbox{$\mathrm{\overline{u}}$} }
\newcommand{\Smeson}{\mbox{$\mathrm{S^{*}/f}_{0}(980)$} }
\newcommand{\dbar}{\mbox{$\mathrm{\overline{d}}$}}
\newcommand{\sbar}{\mbox{$\mathrm{\overline{s}}$}}
\newcommand{\pipi}{\mbox{$\pi^{+}\pi^{-}$} }
\newcommand{\etapi}{\mbox{$\eta\pi$} }
\newcommand{\kpm}{\mbox{K$^{\pm}$} }
\newcommand{\fmeson}{\mbox{$f_{2}(1270)$} }
\newcommand{\pimp}{\mbox{$\pi^{\mp}$} }
\newcommand{\kkpip}{\mbox{$K^{0}_{S} K^{+} \pi^{-}$} }
\newcommand{\kkpim}{\mbox{$K^{0}_{S} K^{-} \pi^{+}$} }
\newcommand{\rhopipi}{\mbox{$\rho^{0}(770)\pi^+\pi^-$} }
\newcommand{\pipipi}{\mbox{$\mathrm{\pi^{+} \pi^{-}}\pi^{0}$} }
\newcommand{\pf}{\mbox{p$_{f}$} }
\newcommand{\ps}{\mbox{p$_{s}$} }
\newcommand{\Om}{\mbox{$\Omega$\ } }
\newcommand{\arr}{\mbox{$\rightarrow$} }
\newcommand{\Apr}{\mbox{\overline{ \rm p} } }
\newcommand{\kzero}{\mbox{$\mathrm{K^{0}$}}}
\newcommand{\piplp}{\mbox{$\mathrm{\pi^{+}p}_{f}$} }
\newcommand{\kkbar}{\mbox{$\mathrm{K\overline{K}}$} }
\newcommand\lsim{\mathrel{\rlap{\lower4pt\hbox{\hskip1pt$\sim$}}
    \raise1pt\hbox{$<$}}}
\newcommand\gsim{\mathrel{\rlap{\lower4pt\hbox{\hskip1pt$\sim$}}
    \raise1pt\hbox{$>$}}}
\newcommand{\ba}{\begin{array}}
\newcommand{\ea}{\end{array}}
\newcommand{\nn}{\nonumber}
\newcommand{\mathbold}{\bf}
\newcommand{\be}{\begin{equation}}
\newcommand{\ee}{\end{equation}}
\newcommand{\bear}{\begin{eqnarray}}
\newcommand{\eear}{\end{eqnarray}}
\newcommand{\tab}{\hspace*{0.5cm}}
\newcommand{\cen}{\hspace*{7.0cm}}
\newcommand{\ii}{\'{\i}}
\newcommand{\II}{\'{\I}}
\newcommand{\sla}{\hspace*{-0.2cm}\slash  }
\newcommand{\slag}{\hspace*{-0.25cm} \slash}

\newcommand{\rvac}{\,|0\rangle}
\newcommand{\lvac}{\langle 0|\,}
\newcommand{\ket}{\,\rangle}
\newcommand{\bra}{\langle \,}
\newcommand{\eqn}[1]{(\ref{#1})}
\newcommand{\cO}{{\cal O}}
\newcommand{\bel}[1]{\be\label{#1}}
\newcommand{\mL}{\mathcal{L}}
\newcommand{\mA}{\mathcal{A}}
\newcommand{\mB}{\mathcal{B}}
\newcommand{\mC}{\mathcal{C}}
\newcommand{\mD}{\mathcal{D}}
\newcommand{\mM}{\mathcal{M}}
\newcommand{\mO}{\mathcal{O}}
\newcommand{\mF}{\mathcal{F}}
\newcommand{\mT}{\mathcal{T}}
\newcommand{\mR}{\mathcal{R}}
\newcommand{\mG}{\mathcal{G}}
\newcommand{\Frac}[2]{\frac{\displaystyle #1}{\displaystyle #2}}
\newcommand{\Int}{\displaystyle{\int}}

\begin{abstract}
We  study the mass, width and couplings  of the lightest 
vector multiplet. 
Effective field theories based on chiral symmetry 
and a $1/N_C$ counting are adopted in order to
describe the vector form factor associated to the two-pseudoscalar
matrix element of the QCD vector current. 
The bare poles of the intermediate $s$--channel 
resonances are regularized
through a Dyson-Schwinger-like summation. This procedure provides
many interesting properties, as the pole mass 
$M_\rho^{\mathrm{pole}}= 764.1\pm 2.7\, {}^{+4.0}_{-2.5}$ MeV 
and the chiral coupling 
$L_9^r(\mu_0)= (7.04\pm 0.05\, {}^{+0.19}_{-0.27})\cdot 10^{-3}$, 
at $\mu_0=770$ MeV. We show how the  running  
affects the resonance parameters and that $\mu$ is really unphysical, 
so saturation occurs at any scale. This talk is mainly based on 
Ref.~\cite{RhoProp}.

\end{abstract}

\maketitle


\section{Introduction}

At high energies ($s\gg 1$ GeV$^2$) Quantum Chromodynamics 
(QCD) admits a perturbative
expansion in powers of the strong coupling cons\-tant. 
Its physical degrees of
freedom are quarks and gluons. However when decreasing the ener\-gy 
($s\lsim 1$ GeV$^2$) the theory
becomes non-perturbative and the asymptotic states are the colourless hadrons. 

In this scenario, effective field theory (EFT) techniques 
are a very powerful tool
which can provide a lot of information about the properties of the hadronic
spectrum. Moreover, the EFTs based on chiral symmetry
have been very successful~\cite{chptms,the role,spin1fields,EC:95}. 
We have focused our attention on 
the study of the vector resonances through the 
pion vector form-factor (VFF). This observable is very sensitive to the
characteristics of these mesons and experiment 
has provided lots of very precise
data~\cite{ALEPH,otherdata,Amendolia}. 
In the isospin limit ($m_u=m_d$) the VFF is
given by  the scalar function $\mF(q^2)$ in  
the hadronic matrix element 
\be
\bra \pi^-\pi^0|\,\bar{d}\gamma^\mu u\rvac\, =\, \sqrt{2} 
\left(p^{\pi^-}-p^{\pi^0}\right)^\mu \, \mF(q^2)\, ,
\ee
being  $q= p^{\pi^-}+p^{\pi^0}$.
  
This talk is focused on the $q^2$--region around the mass
of the re\-so\-nan\-ces. Thus their fields 
have to be explicitly included in the
calculation. We use the quantum field theory (QFT) which handles these fields
preserving chiral invariance: Re\-so\-nan\-ce Chiral Theory
(R$\chi$T)~\cite{the role}.
The calculation of the observable through a well
defined QFT, and in the proper way, implements in an automatic way  
unitarity and analyticity. 

However, since at these energies the chiral counting 
is not valid any longer, we
employed an alternative counting which has resulted very success\-ful in many
former works:  the $1/N_C$ coun\-ting, being $N_C$ the number of colours of QCD
which is considered much larger than one~\cite{NC}.

This work has been a first step to our final aim: 
the calculation of the one loop
quantum corrections in R$\chi$T. In a recent new  work we 
have almost completed a full
regularized one-loop calculation of the VFF~\cite{VFF1loop}.

\section{Leading order}

In the large--$N_C$ limit the VFF is given by tree-level
diagrams, generated by the direct Goldstone coupling from the Chiral
Perturbation Theory ($\chi$PT) Lagrangian
and by the exchange of an infinite tower of resonances~\cite{PI:02}:
\be
\mF \, = \, 1 \, + \, \sum_{i=1}^{\infty} \Frac{F_{V_i} G_{V_i}}{F^2} 
\Frac{q^2}{M_{V_i}^2 -q^2} \, .
\ee
The QCD short-distance behaviour of the 
VFF~\cite{Brodsky}  constrains the couplings at 
 leading order in $1/N_C$ to satisfy \cite{spin1fields,PI:02}
\be
1 \, - \,  \sum_i \Frac{F_{V_i} G_{V_i}}{F^2} \, = \, 0 \, . 
\ee 

At energies below the higher multiplets of resonances 
($\sqrt{s}\lsim 1.2$ GeV), we only need to consider the
lowest-mass $\rho$ state. In that case, the VFF is completely
determined  by the $\rho$ mass:
\bel{eq.VFFLO1res}
\mF \, = \, 1 \, + \,  
\Frac{q^2}{M_{V}^2 -q^2} \, .
\ee  

Such a simple formula provides a spectacular description of the data
\cite{RhoProp,GyP}. 
Of course it fails around the mass of
the $\rho(770)$, since we have a tree-level propagator. This 
is a common problem in perturbation theory: when the energy of an internal 
tree-level propagator becomes on-shell the perturbative expansion fails 
and an all-order Dyson-Schwinger summation must be performed. 
This will be done in the next section, but taking into account 
some peculiarities of the $1/N_C$ counting.

When the energy is increased, Eq.~\eqn{eq.VFFLO1res}  
starts fai\-ling and the inclusion of the next multiplet becomes 
then necessary:
$$
\mF = 1\, + \,\Frac{F_{V_1} G_{V_1}}{F^2} 
\Frac{q^2}{M_{V_1}^2 -q^2}
\, +\, \Frac{F_{V_2} G_{V_2}}{F^2} 
\Frac{q^2}{M_{V_2}^2 -q^2} \, ,
$$  
with the constraint \ $F_{V_1} G_{V_1} + F_{V_2} G_{V_2} = F^2$.   
However, for the energies we are going to consider, the single 
resonance approximation provides a very accurate description.

\section{Dyson-Schwinger summation}

To regularize the real resonance pole an all-order summation is performed, 
having a series of self-energies in the usual way. 
At leading order (LO) in $1/N_C$, the $\rho$ self-energy
is provided by a loop with two pseudoscalar propagators (pions and kaons). 
Therefore, the LO VFF is regulated by the 
re-scattering of the pseudoscalars, which occurs
through intermediate $s$--channel vector resonances and 
through local vertices from the $\cO(p^2)$ $\chi$PT Lagrangian~\cite{Jorge}. 
This yields:
\bel{eq.Dyson}
\mF\, =\,  \Frac{
1 \, + \, \sum_i \Frac{F_{V_i}G_{V_i}}{F^2} 
\Frac{q^2}{M_{V_i}^2-q^2}               
}{  
1 + \Frac{2q^2}{F^2}\left[1+\sum_i\Frac{2G_{V_i}^2}{F^2} 
\Frac{q^2}{M_{V_i}^2-q^2}\right] B_{22}                    
}  \, ,
\ee
with $B_{22}=B_{22}^{(\pi)}+ \frac12 B_{22}^{(K)}$, 
being the Feynman integral
$\int \frac{d^d k}{i(2\pi)^d} \frac{k^\mu k^\nu}{(k^2-m_P^2)((q-k)^2-m_P^2)} 
\equiv B_{22}^{(P)} g^{\mu\nu} q^2 + B_{21}^{(P)} q^\mu q^\nu $ 
from Ref.~\cite{RhoProp}.

The width that arises in the denominator of Eq.~\eqn{eq.Dyson} 
solves the problem of the real pole, and 
this expression will be our final formula for the VFF. 
Nonetheless the real part of the Feynman integral is divergent
and needs some regularization.
How to construct a well defined renormalizable 
theory including resonance fields is still an open problem
(a proper one-loop analysis is going to be finished 
soon~\cite{VFF1loop}). 
Nevertheless, in the next section we will show the way to fix the 
local indetermination associated with this divergence.

\section{Low-energy matching}

At very low energies ($s\ll M_\rho^2$), the Goldstone bosons (pions, 
kaons and eta) are the only relevant degrees of freedom. 
The EFT of QCD in that regime is provided by
$\chi$PT, which is a theory renormalizable order by order.
Performing a matching between our Dyson 
expression~\eqn{eq.Dyson} and the VFF in $\chi$PT,
the ``unknown'' local divergence can be determined.

When making the transition from R$\chi$T to $\chi$PT, 
one must take into account that at LO in 
$1/N_C$ the low-energy $\cO(p^4)$ $\chi$PT couplings 
are completely saturated by the exchange 
of heavy re\-so\-nan\-ces (in the antisymmetric 
formalism)~\cite{the role,spin1fields}. 
However not much is known at NLO, so they still remain 
as free parameters in the theory:
\bel{eq.satuL9}
L_9^r(\mu) \, = \, \underbrace{\Frac{F_V G_V}{2 M_V^2} }_{\cO(N_C)} \, + \, 
\underbrace{\Delta L_9(\mu)}_{\cO(1)}  \, , 
\ee   
where the first and second terms correspond, respectively, 
to the LO and NLO parts of the $\chi$PT  coupling $L_9^r(\mu)$.  

Expanding up to $\cO(q^2)$ 
our result in the one-resonance approximation, we get:
\bel{eq.Dysonop4}
\mF \,  = \, 1 \, + \, \Frac{F_V G_V}{2 M_V^2} \Frac{ 2 q^2}{F^2} 
\, - \, \Frac{2 q^2}{F^2} B_{22} \, + \, \cO(q^4) \, .
\ee   
The corresponding $\chi$PT expression at NLO is:
\bel{eq.CHPTop4}
\mF \,  =  \,  1 \, + \, L_9^r(\mu)
  \Frac{ 2 q^2}{F^2}\, - \, \Frac{2 q^2}{F^2} B_{22}^{\overline{MS}} 
\, \, + \, \cO(q^4) \, ,
\ee
with $L_9^r(\mu)$ related to R$\chi$T by Eq.~\eqn{eq.satuL9} and 
$B_{22}^{\overline{MS}}$ being the 
regularized two-propagator Feynman integral after 
the renormalization of $L_9^r(\mu)$ in the usual $\overline{MS}-1$ 
scheme of $\chi$PT.

The matching of Eqs.~\eqn{eq.Dysonop4}~and~\eqn{eq.CHPTop4} determines 
the Feynman integral $B_{22}$, which becomes dependent on the NLO 
parameter $\Delta L_9(\mu)$:  
$B_{22}\to B^r_{22}=B^r_{22}(\mu, \Delta L_9)$. 

In the now finite Dyson expression, Eq.~\eqn{eq.Dyson}, 
we have both the LO parameter $M_V^2$ and  
$\Delta L_9(\mu)$, but the VFF only \-really\- depends on a very 
specific combination of them which is defined as a running mass:
\be
{M^r_V}(\mu)^2 \,\equiv\, M_V^2 \;
\left[ 1- \frac{2 M_V^2}{F^2} \,\Delta L_9(\mu) \right]\, .
\ee
Substituting $M_V$ in terms of the other two constants,  
the explicit dependence on the NLO parameter 
$\Delta L_9(\mu)$ disappears from the VFF. 
The Feynman integral $B^r_{22}$ becomes then scale dependent, 
being this dependence 
compensated  by the one in the new running mass ${M^r_V}(\mu)^2$. 

Thus, at NLO the resonance couplings have different
values at di\-ffe\-rent scales. 
Moreover, this scale dependence allows us to recover
the full $\chi$PT coupling $L_9^r(\mu)$ 
at any value of $\mu$:
\bel{eq.L9NLO}
L_9^r(\mu)\, = \, L_9^{N_C\to\infty} +\,\Delta L_9(\mu) \, = \, 
\Frac{F_V G_V}{2 {M^r_V}(\mu)^2} \, .
\ee
The importance of this expression is that now there is no 
preferred scale for the resonance sa\-tu\-ra\-tion 
(usually assumed at $\mu=M_\rho$). We have 
obtained a general relation valid at all scales.


\section{Phenomenology}

Using our expression in Eq.~\eqn{eq.Dyson}, we have analyzed
the experimental VFF data from ALEPH~\cite{ALEPH}.
Additional data from other experiments
are also available~\cite{otherdata,Amendolia}.
We firstly made a fit of the region $4m_\pi^2<q^2< (1.2~\mbox{GeV})^2$,
with just one vector multiplet, 
obtaining the re\-so\-nan\-ce couplings 
$F_V$, $G_V$ and ${M_V^r}(\mu)^2$, with a  
good $\chi^2/$dof $=24.8/25$. The fitted VFF, shown in Fig.~\ref{fig.VFFNLO},
is completely insensitive to the chosen value of $\mu$.

The analysis was repeated again including as well the $\rho(1450)$ resonance. 
One gets a similar result, but the $\chi^2/$dof $=14.7/24$
is too small and the behaviour outside the fitted region is worse.
Thus, we took the one-resonance analysis as our best estimate of the VFF.

The parameters obtained from the fit are described 
with more detail in Ref.~\cite{RhoProp}. However,
all these quantities are scale and scheme dependent. 
Therefore we calculated the $\rho(770)$ mass and width 
in two of the more ``physical'' 
scale-independent definitions, 
the Breit-Wigner one~\cite{Sakurai}  and the pole position 
in the complex plane,   
$s_\rho^{\mathrm{pole}}=(M_\rho^{\mathrm{pole}}-
i\,\Gamma_\rho^{\mathrm{pole}}/2)^2$:
%
%
\begin{figure}[tb!]
\hspace*{-0.4cm}
\includegraphics[angle=0,height=6cm,width=0.5\textwidth,clip]{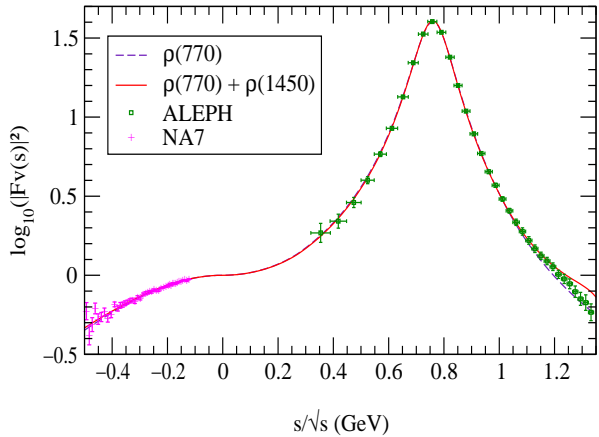}
\vspace*{-1.4cm}
\caption[]{\label{fig.VFFNLO} VFF fits for one and two resonances 
to the ALEPH data~\cite{ALEPH}.
Also shown are the space-like data from NA7~\cite{Amendolia}.  
}
\vspace*{-0.7cm}
\end{figure}
%
%
\be
\ba{l}
M_\rho^{\mathrm{pole}}\, =\, \left(
764.1\pm 2.7\, {}^{+4.0}_{-2.5}\right)\: \mbox{MeV}\, , 
\\[10pt] 
\Gamma_\rho^{\mathrm{pole}}\, =\, \left(
148.2 \pm 1.9\, {}^{+1.7}_{-5.0}\right)\: \mbox{MeV}\, . 
\ea
\ee

\begin{figure}[tb!]
\hspace*{-0.4cm}
\includegraphics[angle=0,height=6cm,width=0.5\textwidth,clip]{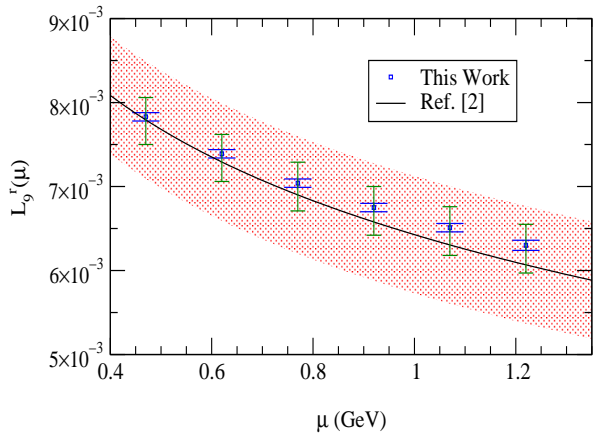}
\vspace*{-1.4cm}
\caption[]{\label{fig.L9run} Variation of $L_9^r(\mu)$ with $\mu$, compared 
with the predicted $\chi$PT running~\cite{chptms}.  
We show statistical and statistical + systematic error bands.  
}
\vspace*{-0.7cm}
\end{figure}

We have also analyzed the influence of the chosen
renormalization scale. The heavy resonance saturation of 
the $\chi$PT coupling constant 
$L_9^r(\mu)$ was studied through Eq.~\eqn{eq.L9NLO}, 
together with the determination of ${M_V^r}(\mu)^2$,
from fits performed at different values of $\mu$. 
The fit was 
repeated for a wide range of scales between 500 and 1200 MeV, 
obtaining the points shown in Fig.~\ref{fig.L9run}.
The agreement with the predicted $\chi$PT  
running~\cite{chptms} is complete, i.e. the high-energy parameters run in the 
proper way to reproduce the low-energy dynamics.
At $\mu_0=770$ MeV, we obtain
$L_9^r(\mu_0)=(7.04\pm 0.05\, {}^{+0.19}_{-0.27})\cdot 10^{-3}$.        
   
\section{Conclusions}

It is important to incorporate the proper energy dependence of the 
observables when making accurate determinations of the resonance and 
chiral parameters. 
This becomes particularly relevant when one tries to obtain
scale-independent properties 
as the pole mass and width, defined in the complex plane, from extrapolations
of the data which sits in the real $q^2$--axis. 
Our analysis 
controls properly the momentum dependence, taking special care of 
ana\-ly\-ti\-ci\-ty and unitarity.

Working within the single-resonance approximation, we have obtained
a good fit to the ALEPH data~\cite{ALEPH}, in the range
$2m_\pi^2<\sqrt{q^2}< 1.2~\mbox{GeV}$. The fitted VFF is not sensitive
to the chosen renormalization scale. However, performing
fits at different values of $\mu$, we recover the proper running of
the $\chi$PT coupling $L_9^r(\mu)$ from the
$\mu$--dependent R$\chi$T parameters, showing how resonance saturation occurs 
at arbitrary scales.

In Ref.~\cite{RhoProp} it was also checked that the deviations of 
the fitted couplings from
their large-$N_C$ predictions~\cite{spin1fields} 
are of the expected $1/N_C$ order, 
i.e. $F^r_V/F=\sqrt{2}+\cO(1/N_C)$ and 
$2\, G^r_V/F=\sqrt{2}+\cO(1/N_C)$. The numerical impact
of the $\rho(1450)$
and of the exchange of heavy resonances in the $t$--channel
was also tested. These effects are tiny for 
$\sqrt{s} \lsim 1.2$~GeV but become relevant at higher energies.

\section*{Acknowledgements}

We thank Stephan Narison for the organization of the QCD 03 Conference 
and Jorge Portol\'es for his many helpful comments.
This work has been supported in part by TMR EURIDICE, EC Contract
No. HPRN-CT-2002-00311, 
by MCYT (Spain) under grant FPA2001-3031 and by ERDF
funds from the European Commission.

\end{document}